\documentclass[aps,pra,twocolumn,amssymb,amsfonts,showpacs,floatfix]{revtex4}
\usepackage[none]{hyphenat}
\usepackage{bm}
\usepackage{dcolumn}% Align table columns on decimal point
\usepackage{bm}% bold math
\usepackage{mathbbol}
\usepackage{graphicx}
\usepackage{epstopdf} 
\usepackage{amsmath,amsthm}
\usepackage{times}
\usepackage{microtype}
\usepackage{color}
\usepackage{titlesec}
\titlespacing\section{0pt}{12pt plus 4pt minus 2pt}{2pt plus 2pt minus 2pt}
\titlespacing\subsection{0pt}{12pt plus 4pt minus 2pt}{2pt plus 2pt minus 2pt}
\titlespacing\subsubsection{0pt}{12pt plus 4pt minus 2pt}{2pt plus 2pt minus 2pt}
\begin{document}
\title{Nonexponential fidelity decay in isolated interacting quantum systems}

\author{E. J. Torres-Herrera}
\author{Lea F. Santos}
\affiliation{Department of Physics, Yeshiva University, New York, New York 10016, USA}

\date{\today}
%\date{Received 17 July 2014; published 22 September 2014}

\begin{abstract}
We study isolated finite interacting quantum systems after an instantaneous perturbation and show three scenarios in which the probability for finding the initial state later in time (fidelity) decays nonexponentially, often all the way to saturation. The decays analyzed involve Gaussian, Bessel of the first kind, and cosine squared functions. The Gaussian behavior emerges in systems with two-body interactions in the limit of strong perturbation. The Bessel function, associated with the evolution under full random matrices, is obtained with surprisingly sparse random matrices. The cosine squared behavior, established by the energy-time uncertainty relation, is approached after a local perturbation in space.
\end{abstract}

%03.65.-w 	Quantum mechanics
%75.10.Pq 	Spin chain models
%75.10.Jm 	Quantized spin models, including quantum spin frustration
%03.65.Xp 	Tunneling, traversal time, quantum Zeno dynamics
% 05.45.Mt 	Quantum chaos; semiclassical methods

\maketitle

%%%%%%%%%%%%%%%% INTRODUCTION %%%%%%%%%%%%%%%%%%%%%%%%%
\section{INTRODUCTION}
The time evolution of isolated quantum systems out of equilibrium has been explored for many decades. The subject is strongly connected with the derivation of the energy-time uncertainty relation~\cite{Mandelstam1945,Ersak1969,Fleming1973,Bhattacharyya1983,Gislason1985,Vaidman1992,Uffink1993,Pfeifer1993,GiovannettiPRA2003,Boykin2007}, since the lifetime of a decaying state is bounded by the reciprocal of the energy uncertainty.  It has been central to studies of unstable systems, sometimes in relationship with quantum chaos and notions of quantum ergodicity~\cite{Weisskopf1930,Khalfin1958,Fonda1978,Peres1984,Greenland1988,Wilkinson1997,Flambaum2000A,Flambaum2001a,Flambaum2001b,Izrailev2006,Jacquod2001,Cucchietti2002,Benenti2002,Prosen2002,Wisniacki2003,Rothe2006,Gorin2004,Gorin2006,Ng2006,Campo2011,Benet2011,Wisniacki2013,Fine2014}. It is at the heart of progresses in quantum information and the development of methods to control the dynamics of quantum systems~\cite{Cappellaro2007,Ramanathan2011,Jalabert2001,Emerson2002,Weinstein2002,Jacquod2009,Caneva2009,Goussev2012,Zangara2012}. Recently, it has become an important topic for experiments in optical lattices, where many-body quantum systems can evolve coherently for long times~\cite{Trotzky2008,Chen2011,Trotzky2012,Fukuhara2013}, and related theoretical studies~\cite{Silva2008,Silva2009,Genway2010,Venuti2011,Gramsch2012,Heyl2012,HeSantos2013,Shchadilova2014,Andraschko2014,Schiro2014}.

Here, we analyze the probability to find an isolated interacting quantum system in its initial state later in time. This probability, often called nondecay probability, return probability, or survival probability, is denominated here as fidelity. The picture considered is that of an instantaneous quench, where the system is initially in an eigenstate of an initial Hamiltonian $\widehat{H}_I$ and the dynamics is launched by changing $\widehat{H}_I$ abruptly into a new final Hamiltonian $\widehat{H}_F$.  The fidelity is  obtained by Fourier transforming the weighted energy distribution of the initial state. This distribution is known as the local density of states (LDOS) or strength function.

The fidelity decay is exponential when the LDOS has a Lorentzian (also known as Cauchy or Breit-Wigner) form. This is the common behavior in open systems~\cite{Fonda1978}, although algebraic contributions at long times have been predicted as early as 1958 \cite{Khalfin1958}. They are caused by the lower cutoff in the energy distribution. The purpose of the present paper is to show that in isolated interacting quantum systems, various deviations from the Lorentzian shape may occur. We present three realistic cases that lead to nonexponential decays. The paradigmatic interacting quantum systems used for the illustrations are one-dimensional spin-1/2 models.

{\em Case (1)}. In previous works~\cite{Torres2014PRA,Torres2014NJP,Torres2014PRE,TorresKollmarARXIV}, we emphasized that in the limit of strong global perturbation, the LDOS becomes Gaussian, which causes a Gaussian fidelity decay. Such behavior, even at long times, had been discussed before~\cite{Flambaum2000A,Flambaum2001a,Flambaum2001b,Wisniacki2003,Izrailev2006}. We showed that it can in fact persist all the way to saturation independently of the regime (integrable or chaotic) of the system. We now extend those analyses, concentrating on the transition region from Lorentzian to Gaussian and on the deformations that the Gaussian distribution undergoes as the energy of the initial state moves away from the center of the spectrum of $\widehat{H}_F$. 

The two other cases explored give rise to fidelity decays that are even faster than Gaussian.

{\em Case (2)}. The matrix elements of the final Hamiltonian associated with the spin flip-flop terms (excitation hopping) between second and further neighbors are randomized. This, at first sight, very sparse random matrix leads to a semicircular LDOS and to a fidelity behavior involving a Bessel function of the first kind. This decay is very similar to that found when $\widehat{H}_F$ is a full random matrix.

{\em Case (3)}. Contrary to the cases above, where the LDOS is unimodal, the third scenario corresponds to a bimodal distribution where  the two peaks are far in energy. The initial fidelity decay reaches the quantum limit as established by the energy-time uncertainty relation, following a cosine squared function. The two peaks are created by adding to the initial spin-1/2 Hamiltonian a local and very strong static magnetic field that splits the spectrum of the final Hamiltonian in two.

The article is organized as follows. The model and quenches considered are described in Sec.~\ref{Sec:model}. Section~\ref{Sec:fidelity} explains the relationship between the fidelity decay and the LDOS. The three nonexponential decays are studied in Secs.~\ref{Sec:Gaussian}, \ref{Sec:FRM}, and \ref{Sec:Cos}. Final remarks are presented in Sec.~\ref{Sec:Conclusion}.

%%%%%%%%%%%%%%%% MODEL and QUENCH %%%%%%%%%%%%%%%%%%%%%%%%%
\section{MODEL AND QUENCH}
\label{Sec:model}
We consider one-dimensional spin-1/2 lattices with two-body interactions. They are used to model quantum computers, real magnetic compounds, and nuclear magnetic resonance systems and are currently being investigated with cold atoms in optical lattices. The Hamiltonian for $L$ sites and open boundary conditions is given by
\begin{equation}
\widehat{H} =  d J \widehat{S}_{ L/2 }^z + \widehat{H}_{\text{NN}}+ \lambda \widehat{H}_{\text{NNN}},
\label{eq:Ham}
\end{equation}
where
\begin{eqnarray}
&&\widehat{H}_{\text{NN}} = J\sum_{k=1}^{L-1} \left( 
\widehat{S}_k^x \widehat{S}_{k+1}^x + \widehat{S}_k^y \widehat{S}_{k+1}^y +
\Delta \widehat{S}_k^z \widehat{S}_{k+1}^z \right),  
\nonumber \\
&&\widehat{H}_{\text{NNN}} = J\sum_{k=1}^{L-2} \left(\widehat{S}_k^x \widehat{S}_{k+2}^x + \widehat{S}_k^y \widehat{S}_{k+2}^y
+ \Delta \widehat{S}_k^z \widehat{S}_{k+2}^z  \right). \nonumber
\end{eqnarray}
Above, $\hbar =1$ and $\widehat{S}^{x,y,z}_k $ are spin operators acting on site $k$. $\widehat{S}_k^x \widehat{S}_{k+1}^x + \widehat{S}_k^y \widehat{S}_{k+1}^y$ $[\widehat{S}_k^x \widehat{S}_{k+2}^x + \widehat{S}_k^y \widehat{S}_{k+2}^y]$ is the flip-flop term and $\widehat{S}_k^z \widehat{S}_{k+1}^z [\widehat{S}_k^z \widehat{S}_{k+2}^z]$ is the Ising interaction between nearest-neighbor (NN) [next-nearest neighbor (NNN)] spins. $J$ is the exchange coupling constant. In what follows, $J=1$ sets the energy scale. $\Delta$ is the anisotropy parameter and $\lambda$ refers to the ratio between NNN and NN couplings. These two parameters are assumed positive and $L$ is chosen to be even.

The total spin in the $z$ direction, $\widehat{{\cal{S}}}^z=\sum_i\widehat{S}_i^z$, is conserved. We deal with subspaces of dimension ${\cal D}=L!/[(L-u)! u!]$, where $u$ is the number of spins pointing up in $z$. Other symmetries include spin reversal, when ${\cal{S}}^z=0$; conservation of total spin, when $\Delta=1$; and parity, when the impurity term $d J \widehat{S}_{ L/2 }^z$ in the middle of the chain is not present. This term leads to a Zeeman splitting that is different from that on the other sites. 

The system with only NN couplings is integrable~\cite{Bethe1931}. The impurity~\cite{Santos2004,Gubin2012,Torres2014PRE} or the inclusion of NNN couplings ~\cite{Hsu1993,Gubin2012,Kudo2005} can take the system into the chaotic regime.

We refer to Hamiltonian (\ref{eq:Ham}) as 

(i) XX when $d, \Delta, \lambda=0$; 

(ii) XXZ when $d, \lambda=0$ and $\Delta \neq 0$; 

(iii) Impurity model when $d, \Delta\neq0$ and $\lambda=0$; 

(iv) NNN model when $d=0$ and $\Delta, \lambda \neq0$.

\subsection*{Quench dynamics} 

The system starts in an excited eigenstate, $|\Psi(0)\rangle =|\text{ini}\rangle$, of an initial Hamiltonian $\widehat{H}_I$. In most cases studied below, $\widehat{H}_I$ is the XXZ model. This is the integrable part of $\widehat{H}$ (\ref{eq:Ham}) and corresponds to the mean-field unperturbed Hamiltonian.

After an instantaneous perturbation, the state evolves as
\begin{equation}
|\Psi(t)\rangle = e^{-i \widehat{H}_\text{F} t} |\text{ini} \rangle=
\sum_{\alpha} C_{\alpha}^{\text{ini}} e^{-i E_{\alpha} t} |\psi_{\alpha}  \rangle ,
\end{equation}
where $E_{\alpha} $ and $|\psi_{\alpha}\rangle$ are the eigenvalues and eigenstates of the final Hamiltonian $\widehat{H}_F$ and $C_{\alpha}^{\text{ini}} = \langle \psi_{\alpha} |\text{ini} \rangle$. The eigenvalues and eigenstates of the initial Hamiltonian are denoted by ${\cal E}_n$ and $|n\rangle $. The subscripts ``$I$" and ``$F$" are used for the Hamiltonians and also for their parameters before and after the quench, respectively.

%%%%%%%%%%%%%%%% LDOS and FIDELITY %%%%%%%%%%%%%%%%%%%%%%%%%
\section{FIDELITY}
\label{Sec:fidelity}

The quantum fidelity measures how close two quantum states are. In the case of two pure states, it is defined as the absolute squared value of the overlap between them. Here, we study the fidelity between the initial state and its evolved counterpart,
\begin{equation}
F(t) =\left| \langle \text{ini} | e^{-i \widehat{H}_\text{F} t} |\text{ini} \rangle \right|^2  = \left|\sum_{\alpha} |C_{\alpha}^{\text{ini}} |^2 e^{-i E_{\alpha} t}  \right|^2.
\label{eq:fidelity}
\end{equation}
In this case, $F(t)$ coincides with the survival probability. It measures the probability for finding the initial state later in time; that is, it quantifies the level of stability of the quantum system. From the equation above, one sees that the fidelity is simply the Fourier transform in energy of the components $|C_{\alpha}^{\text{ini}} |^2$.

The distribution of $|C_\alpha^\text{ini}|^2$ in the eigenvalues $E_{\alpha}$,
\begin{equation}
P^{\text{ini}}(E) = \sum_\alpha |C_\alpha^\text{ini}|^2\delta(E-E_\alpha),
\end{equation}
is the LDOS and it is also related to the work distribution function~\cite{Silva2008}.
When ${\cal D}$ is large and the LDOS is dense,  the sum in Eq.~(\ref{eq:fidelity}) can be substituted by an integral,
\begin{equation}
F(t) \approx \left|  \int_{-\infty}^{\infty} P^{\text{ini}}(E) e^{-i E t} dE \right|^2,
\end{equation}
where $P^{\text{ini}}(E)$ is now the envelope of the LDOS. In spectroscopy, $P^{\text{ini}}(E)$ is the spectral line shape and its characteristic function is the time-domain signal.

Very often in spectroscopy and also in nuclear and particle physics, the fidelity decays exponentially. This is a distinctive feature of unstable systems and implies a Lorentzian line shape,
\begin{eqnarray}
&&P^{\text{ini}}_{\text{L}}(E) = \frac{1}{2\pi} \frac{\Gamma_{\text{ini}} }{(E_{\text{ini}} - E)^2 +
 \Gamma_{\text{ini}}^2 /4} ,  \nonumber  \\[1em]
 &&\Rightarrow  F_{\text{L}}(t)=\exp (-\Gamma_{\text{ini}} t), 
 \label{eq:BW}
\end{eqnarray}
where $\Gamma_{\text{ini}} $ is the full width at half maximum of the distribution. However, deviations from the exponential behavior at short and at long times have been discussed very early in the studies of unstable quantum systems~\cite{Khalfin1958,Fonda1978,Greenland1988}. 

Power-law decays at long times were examined in continuous spectra bounded from below~\cite{Khalfin1958,Fonda1978,Greenland1988,Rothe2006,Campo2011}. It has been observed also in systems at the Anderson metal-insulator transition~\cite{Ketzmerick1992,Ketzmerick1997}.
At very short times, the expected behavior is quadratic in $t$.  The Taylor expansion of $e^{-i E_{\alpha} t} $ in Eq.~(\ref{eq:fidelity}) leads to
\begin{equation}
F(t) \approx 1-\sigma_\text{ini}^2t^2,
\label{eq:fidelity_approx}
\end{equation}
where
\begin{equation}
\sigma_{\text{ini}} = \sqrt{\sum_{\alpha} |C_{\alpha}^{\text{ini}} |^2 (E_{\alpha} - E_{\text{ini}})^2}
=\sqrt{\sum_{n \neq \text{ini}} |\langle n |\widehat{H}_F | \text{ini}\rangle |^2 }
\label{deltaE}
\end{equation}
is the energy dispersion of $|\text{ini}\rangle $ and 
\begin{equation}
E_{\text{ini}} = \langle \text{ini} |\widehat{H}_F | \text{ini} \rangle = \sum_{\alpha} |C_{\alpha}^{\text{ini}}|^2 E_{\alpha} 
\label{Eini}
\end{equation}
is the energy of the initial state projected on the final Hamiltonian. 
Clearly, the short-time behavior shown in Eq.~(\ref{eq:fidelity_approx}) cannot be achieved by expanding the exponential expression in Eq.~(\ref{eq:BW}). As matter of fact, $\sigma_{\text{ini}}$ is infinite for the Lorentzian function, which forces the energy-time uncertainty relation in systems with exponential fidelity decays to be written in terms of $\Gamma_{\text{ini}}$ instead of $\sigma_{\text{ini}}$ \cite{Uffink1993}. 

The Lorentzian shape for $P^{\text{ini}}(E)$ is not universal. It emerges under the assumption that the initial state is coupled to infinitely many states with coupling strengths of the same order~\cite{BohrBook,ZelevinskyRep1996}. It is a very good approximation, in agreement with observed exponential decays, when the couplings with the initial state are nonperturbative, although not very strong~\cite{Flambaum2000}. But deviations do exist.

In the limit of strong perturbation, the LDOS for isolated systems with two-body interactions becomes Gaussian~\cite{ZeleB,ZelevinskyRep1996,Flambaum1997,Flambaum2000,Flambaum2000A,Flambaum2001a,Flambaum2001b,Kota2001PRE,Chavda,KotaBook,Santos2012PRL,Santos2012PRE,Torres2013}, causing the Gaussian fidelity decay,
\begin{eqnarray}
&&P^{\text{ini}}_{\text{G}}(E) = \frac{1}{ \sqrt{ 2 \pi \sigma^2_{\text{ini}} }  } \exp \left[ -\frac{(E-E_{\text{ini}})^2}{ 2 \sigma^2_{\text{ini}} }   \right], 
\nonumber \\ [1em]
&&\Rightarrow F_{\text{G}}(t)=\exp(-\sigma_\text{ini}^2t^2) ,
\label{Gauss_SF} 
\end{eqnarray}
which agrees with Eq.~(\ref{eq:fidelity_approx}) at short times. The Gaussian behavior was expected to hold for some time and then switch to exponential at longer times. We have shown several cases, some accessible to experiments in optical lattices, where $F(t)$ can in fact be Gaussian all the way to saturation~\cite{Torres2014PRA,Torres2014NJP,Torres2014PRE,TorresKollmarARXIV}. In Sec.~\ref{Sec:Gaussian} we expand this analysis.

The fact that $P^{\text{ini}}(E) $ can become Gaussian is a reflection of the density of states of systems with two-body interactions, which is also Gaussian~\cite{French1970,Bohigas1971,Brody1981}. In such systems, the maximum possible spreading of the LDOS is given by the Gaussian envelope in Eq.~(\ref{Gauss_SF}), which is known as the energy shell. 

If the density of states of $\widehat{H}_F$ is other than Gaussian, we may find $P^{\text{ini}}(E) $ leading to faster than Gaussian fidelity decays. When $P^{\text{ini}}(E) $ is unimodal, the lower bound for $F(t)$  is achieved when $\widehat{H}_F$ is a full random matrix. In this case, the density of states is semicircular, as derived by Wigner~\cite{Wigner1955,Wigner1957,PorterBook,Guhr1998}, and so is the LDOS~\cite{Torres2014PRA,Torres2014NJP},
\begin{eqnarray}
&&P^{\text{ini}}_{\text{SC}}(E)= \frac{1}{\pi \sigma_{\text{ini}} } \sqrt{1 - \left(\frac{E}{2 \sigma_{\text{ini}} }\right)^2}, 
\nonumber \\[1em]
&&\Rightarrow F_{\text{SC}}(t)=\frac{ [{\cal J}_1( 2 \sigma_{\text{ini}} t)]^2}{\sigma_{\text{ini}}^2 t^2} ,
\label{SemiCirc_SF} 
\end{eqnarray}
where $4 \sigma_{\text{ini}} $ is the length of the spectrum and ${\cal J}_1$ is the Bessel function of the first kind. Notice that $F_{\text{SC}}(t)$ also agrees with Eq.~(\ref{eq:fidelity_approx}) at short times.  In Sec.~\ref{Sec:FRM} we look for models more plausible than full random matrices where the fidelity decay approaches $F_{\text{SC}}(t)$.

The ultimate bound for the fidelity decay, as derived from the energy-time uncertainty relation~\cite{Mandelstam1945,Fleming1973,Bhattacharyya1983}, is given by $ F(t)\geq \cos^2 (\sigma_{\text{ini}} t)$. It is valid for $0\leq t \leq \pi/(2 \sigma_{\text{ini}})$ and agrees with Eq.~(\ref{eq:fidelity_approx}) at short times. This bound can be reached when $P^{\text{ini}}(E) $ is bimodal,
\begin{eqnarray}
&&P^{\text{ini}}_{\text{C}}(E)= \frac{\delta(E_1) + \delta(E_2)}{2} , 
\nonumber \\[1em]
&&\Rightarrow F_{\text{C}}(t)=\cos^2 \left[\frac{(E_2-E_1)t}{2}\right] .
\label{Cos_SF} 
\end{eqnarray}
In Sec.~\ref{Sec:Cos}, we explore a more realistic situation, where $P^{\text{ini}}(E) $ has two non-$\delta$-function peaks and the initial decay is indeed described by $F_{\text{C}}(t)$.

\subsection*{Fidelity decay saturation}

The systems studied here are finite, so after a dephasing time, the fidelity saturates and simply fluctuates around its infinite time average, $\overline{F}=\sum_{\alpha} |C_{\alpha}^{\text{ini}} |^4$. The fluctuations around the saturation point do not die out completely, but they decrease with system size~\cite{Zangara2013,Torres2014NJP,TorresKollmarARXIV}. 

We denote by $t_R$ the time that it takes for the fidelity to first reach $\overline{F}$. When the LDOS is dense and unimodal, as in Secs.~\ref{Sec:Gaussian} and \ref{Sec:FRM}, the difference between the dephasing time and $t_R$ is small, but when $P^{\text{ini}}(E)$ is bimodal, as in Sec.~\ref{Sec:Cos}, large oscillations can survive for a fairly long time after $t_R$.

%%%%%%%%%%%%%%%% GAUSSIAN %%%%%%%%%%%%%%%%%%%%%%%%%
\section{GAUSSIAN DECAY}
\label{Sec:Gaussian}

We emphasize that fast fidelity decays, such as exponential or Gaussian, are not exclusive to chaotic postquench Hamiltonians. They are found also in integrable systems. The decay rate is determined by the shape of the LDOS not the regime  (integrable or chaotic) of  $\widehat{H}_F$.

\begin{figure}[htpb]
\centering
\includegraphics*[width=3.in]{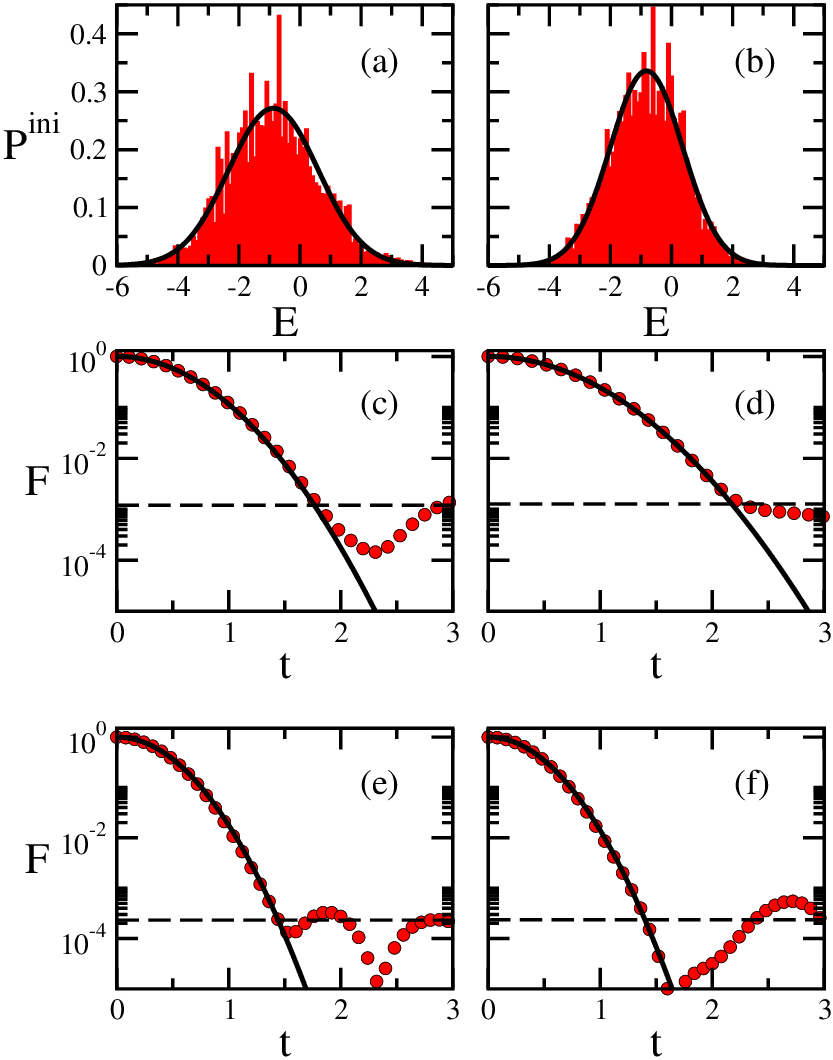}
\caption{(Color online) Local density of states (a, b) and fidelity decay (c-f). Quench from the XX to the XXZ Hamiltonian with $\Delta_F=1.5$, $E_\text{ini}=-0.87$, $\sigma_\text{ini}=1.47$, and $\overline{F}=1.19 \times 10^{-3}$ (a, c). Quench from the XXZ  to the NNN model with $\lambda_F=1$, $\Delta_I=\Delta_F=0.5$, $E_\text{ini}=-0.82$, $\sigma_\text{ini}=1.17$, $\overline{F}=1.326 \times 10^{-3}$ (b, d).  In panels (e) and (f), the initial state is an eigenstate of a full random matrix from a GOE projected onto the XXZ Hamiltonian, $E_\text{ini}=-0.39$, $\sigma_\text{ini}=2.01$, and $\overline{F}=2.32 \times 10^{-4}$ (e), and onto the NNN model, $E_\text{ini}=-0.24$, $\sigma_\text{ini}=2.07$, $\overline{F}=2.35 \times 10^{-4}$ (f). The solid lines give the analytical Gaussian expressions [Eq.~(\ref{Gauss_SF})], and the shaded area (top panels) and circles (middle and bottom panels) are numerical results. The saturation value of the fidelity is indicated with the dashed horizontal line; $L=16$, $\widehat{\mathcal{S}}^z=0$, and ${\mathcal{D}} = 12\,870$; and $J=1$ sets the energy scale.
}
\label{fig:regime}
\end{figure}

In Fig.~\ref{fig:regime}, we consider quenches in the limit of strong perturbation. In Fig.~\ref{fig:regime}(a), the quench is between integrable Hamiltonians, from the XX model to the XXZ model with $\Delta_F=1.5$ (we avoid $\Delta_F=1$, because this is a critical point), and in Fig.~\ref{fig:regime}(b), the quench is from the integrable XXZ model 

\noindent to the chaotic NNN model with $\lambda_F=1$. They make it evident that the filling of the energy shell  does not depend on the regime of the postquench Hamiltonian, but on the  interplay between the initial state and the final Hamiltonian. The shell can be substantially filled when the final Hamiltonian is chaotic and  also when it is integrable, provided $E_{\text{ini}}$ is close to the middle of the spectrum of $\widehat{H}_F$, which is the case in the figures. 

The corresponding fidelity decays for both quenches [Figs.~\ref{fig:regime}(c) and~\ref{fig:regime}(d)] are Gaussian, following Eq.~(\ref{Gauss_SF}). For the chosen parameters, $\sigma_{\text{ini}}$ is actually larger for the integrable-integrable quench than for the integrable-chaotic quench, which explains why the decay in Fig.~\ref{fig:regime}(c) is faster than that in Fig.~\ref{fig:regime}(d). This serves as a good example against the common expectation that fidelity decays should be necessarily faster in chaotic systems.

The Gaussian behavior in Figs.~\ref{fig:regime}(c) and~\ref{fig:regime}(d) holds all the way to saturation. The saturation point, indicated with the horizontal dashed line, is also determined by the interplay between the initial state and the final Hamiltonian. The values of $\overline{F}$  in Figs.~\ref{fig:regime}(c) and~\ref{fig:regime}(d) are very close. In a previous work~\cite{Torres2014NJP}, we found cases where the saturation point for chaotic systems was smaller than that for integrable models and cases where it was even larger. The latter happened when $E_{\text{ini}}$ was further from the middle of the spectrum  for the chaotic model than for the integrable one.

Figures~\ref{fig:regime}(e) and~\ref{fig:regime}(f) show the fidelity decay for the extreme case where the initial state corresponds to a random vector extracted from a full random matrix of an ensemble where the matrices are real and symmetric [Gaussian orthogonal ensemble (GOE)]. The state is then evolved with the same XXZ model [Fig.~\ref{fig:regime}(e)] and the NNN model  [Fig.~\ref{fig:regime}(f)] considered in the top panels. This is an initial state with infinite temperature placed in the middle of the spectrum. The filling of the energy shell is ergodic for both final Hamiltonians. The decay is Gaussian all the way to saturation and evidently faster than that in Figs.~\ref{fig:regime}(c) and~\ref{fig:regime}(d). The saturation point is equivalent for the two systems and given by $\overline{F} \approx 3/{\cal D}$, as expected for normalized random vectors from GOEs.

Below, we expand the Gaussian decay analysis for two other scenarios. First, we study the transition of $P^{\text{ini}}(E) $ from Lorentzian to Gaussian as the perturbation strength increases. We focus on initial states close to the middle of the spectrum. Next, we concentrate on the strong perturbation scenario, but move $E_{\text{ini}}$ away from the middle of the spectrum.

\subsection{Middle of the spectrum}
\label{sec:transition}

When the perturbation is very weak and $\widehat{H}_I \sim \widehat{H}_F$, the LDOS is close to a delta function. As the perturbation increases, the distribution broadens and first becomes Lorentzian. As it increases even further, $P^{\text{ini}}(E) $ eventually reaches the Gaussian shape. Different functions have been used to describe the transition region between Lorentzian and Gaussian. In Refs.~\cite{Chavda,KotaBook}, the Student's t-distribution was employed. It has a Bell shape like the Gaussian distribution, but the tails decrease more slowly. In Ref.~\cite{Flambaum2001a}, an approximate expression combining $P_{\text{L}}(E)$  and $P_{\text{G}}(E) $ to interpolate between the short-time quadratic behavior and the later exponential decay of $F(t)$ was investigated. 

In spectroscopy, emission and absorption lines often have a shape that lies between Lorentzian (when homogeneous broadening dominates) and Gaussian (when inhomogenous broadening is important). They are fitted with the Voigt function, which is a convolution of a Lorentzian and a Gaussian. However, numerical convolutions are computationally expensive. This motivated the introduction of the pseudo-Voigt distribution~\cite{Wertheim1974}, which is simply a linear combination of the two functions with the same full width at half maximum, $ 2 \sqrt{2 \ln 2} \; \sigma_{\text{pV}} =\Gamma_{\text{pV}}$,
\begin{eqnarray}
&&P^{\text{ini}}_{\text{pV}}(E)=\eta P_{\text{L}}(E) + (1-\eta) P_{\text{G}}(E),
 \nonumber \\[1em]
&&\Rightarrow F_{\text{pV}}(t)=\eta^2 \exp (-\Gamma_{\text{pV}} t) + (1-\eta)^2 \exp(-\sigma_{\text{pV}}^2 t^2) \nonumber \\
&& + 2 \eta (1-\eta) \exp \left(-\frac{\Gamma_{\text{pV}} t + \sigma_{\text{pV}}^2 t^2 }{2}\right),
\label{eq:pVoigt} 
\end{eqnarray}
where $0\leq \eta \leq 1$. There are other approximations to the Voigt function~\cite{Thomson1987,Ida2000,Limandri2008}, but this is a simple and fairly good one.

Here, we employ the pseudo-Voigt distribution and its Fourier transform to describe the transition of $P^{\text{ini}}(E) $ from Lorentzian to Gaussian and the changes in the fidelity decay as we increase the strength $\lambda_F$ in the quench from the XXZ to the NNN model. We quantify the transition with the parameter $\eta$ from Eq.~(\ref{eq:pVoigt}) and estimate the critical time $t_c$ for the switch from the Gaussian to the exponential fidelity decay. In the limit of strong perturbation, $\eta \rightarrow 0$ and $t_c \rightarrow t_R$.

\begin{figure}[htb]
\centering
\includegraphics*[width=3.3in]{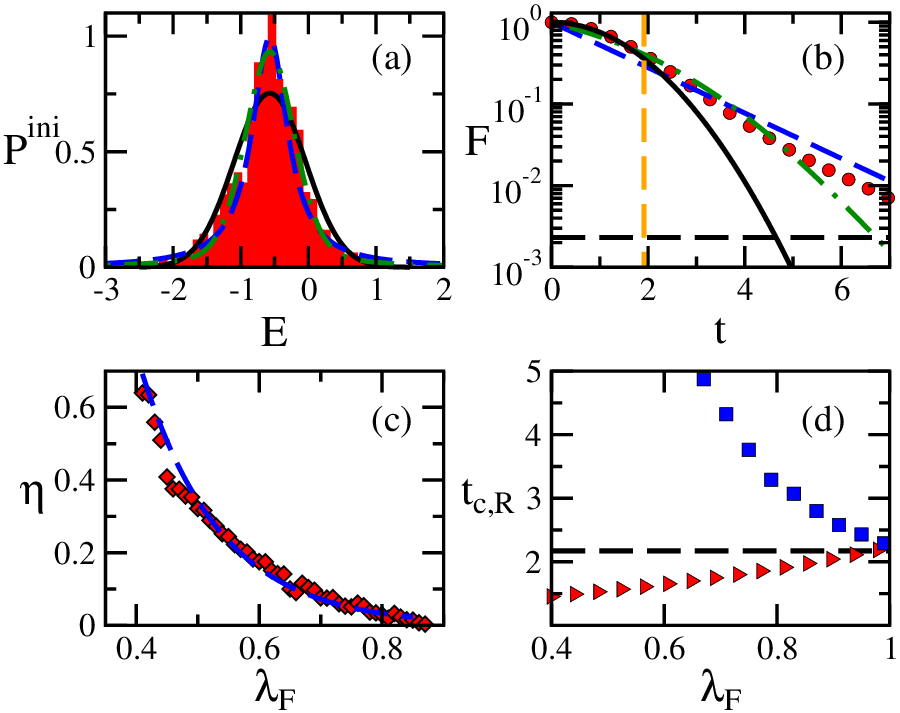}
\caption{(Color online) Quench from the XXZ to the NNN model, $\Delta_I=\Delta_F=0.5$. Local density of states (a) and fidelity decay (b) for $\lambda_F=0.45$. Parameter $\eta$ of the pseudo-Voigt distribution vs $\lambda_\text{F}$ (c); critical time and and $t_R$ vs $\lambda_\text{F}$ (d). The initial state for all panels has ${\cal E}_{5445} =-0.37$. In panels (a) and (b), $E_\text{ini}=-0.57$, $\sigma_\text{ini}=0.53$, and $\overline{F}=2.32\times 10^{-3}$. The fittings lead to $\Gamma_\text{ini}=0.64$, $\sigma_{\text{pV}}=0.37$, and $\eta=0.41$. Numerical results, shaded area and symbols; Gaussian, black solid lines; Lorentzian, blue dashed lines; pseudo-Voigt, green dot-dashed lines; saturation point, horizontal line); and value of $t_c$, vertical line. (c) Fitting for the numerical data (dashed line). (d) $t_c$ (triangles), $t_R$ (squares), and $t_R$ for $\lambda_F=1$ (horizontal dashed line). $L=16$, $\widehat{\mathcal{S}}^z=0$, and ${\mathcal{D}} = 12\,870$; $J=1$ sets the energy scale.}
\label{fig:transitionLG}
\end{figure}

To illustrate the transition region, we use in Fig.~\ref{fig:transitionLG}(a) and~\ref{fig:transitionLG}(b) an intermediate value of the perturbation parameter, $\lambda_F=0.45$. Figure~\ref{fig:transitionLG}(a) shows the LDOS of an initial state close to the middle of the spectrum. All curves consider the same $E_{\text{ini}}$ [Eq.~(\ref{Eini})] calculated from the numerical data. The Gaussian function is the analytical expression from Eq.~(\ref{Gauss_SF}), using the numerical data to compute $\sigma_{\text{ini}}$ [Eq.~(\ref{deltaE})]. The Lorentzian distribution is obtained by fitting $\Gamma_{\text{ini}}$. The pseudo-Voigt function is obtained by fitting $\sigma_{\text{pV}}$ and $\eta$. Figure~\ref{fig:transitionLG}(b) gives the corresponding fidelity curves. The numerical data shows a Gaussian decay for $t<t_c$ and then exponential for $t>t_c$. The approximate value of the critical time is indicated in the figure with the vertical dashed line. $F_{\text{pV}}(t)$ is quite successful in capturing both behaviors, demonstrating that the pseudo-Voigt is a better match to $P^{\text{ini}}(E) $ than the Gaussian or the Lorentzian.

The approach of the LDOS  to the Gaussian shape with the perturbation strength is made evident with Fig.~\ref{fig:transitionLG}(c), where $\eta$ decreases to zero as $\lambda_F$ increases. The numerical data are reasonably well fitted with an exponential function (dashed line).

The dependence of $t_c$ on $\lambda_F$ is shown in Fig.~\ref{fig:transitionLG}(d). The value of the critical time is obtained by finding a local minimum in the vicinity of $t=2$ for the distance between the Gaussian analytical expression and an exponential fitting for $F(t)$. The estimate is rough and very dependent on the time interval used for the exponential fitting. We were able to find the local minimum for $\lambda_F$ up to 0.8, the subsequent points in the figure being an extrapolation. Still, the plot gives a good idea of the increase of $t_c$ with $\lambda_F$ and its approach to $t_R$, indicating that at strong perturbation the decay can indeed be Gaussian all the way to saturation.

\subsection{Away from the middle of the spectrum}

Less attention has been given to the analysis of the shape of the LDOS as $E_{\text{ini}}$ moves away from the center of the spectrum of the final Hamiltonian~\cite{Torres2013}. We study this scenario here for the quench from the XXZ to the NNN model in the limit of strong perturbation, $\lambda_F=1$. 

Since the density of states of systems with two-body interactions is Gaussian, at low energies, the states are fewer and more localized. As the initial state approaches this region, $P^{\text{ini}}(E)$ becomes less dense, slowing down the fidelity decay, and it also becomes more skewed.

\begin{figure}[htb]
\centering
\includegraphics*[width=3.3in]{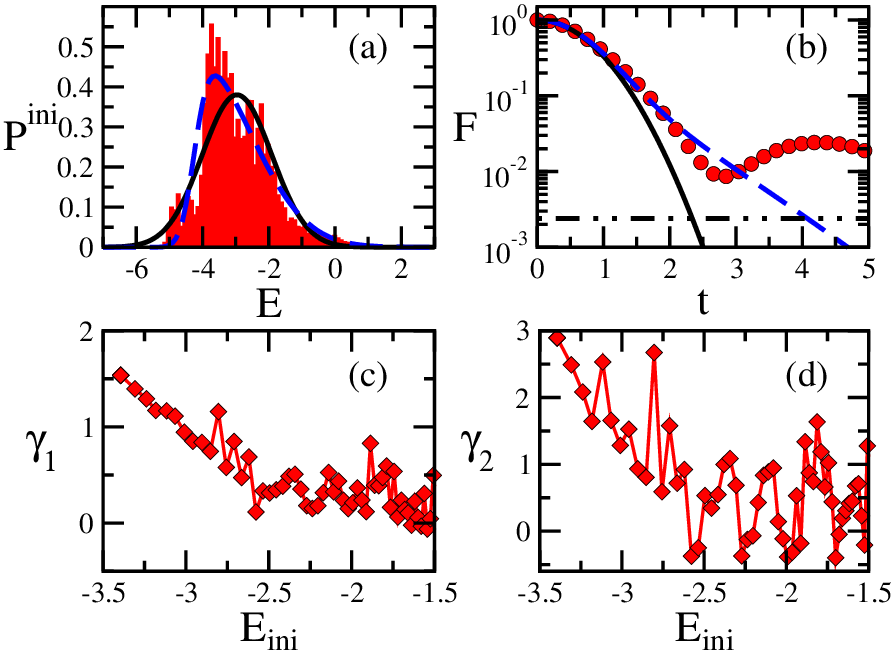}
\caption{(Color online) Quench from the XXZ to the NNN model, $\Delta_I=\Delta_F=0.5$, and $\lambda_F=1$. Local density of states (a) and fidelity decay (b). Skewness $\gamma_1$ (c) and excess kurtosis $\gamma_2$ (d) vs $E_\text{ini}$. In panels (a) and (b), $E_\text{ini}=-2.96$, $\sigma_\text{ini}=1.05$, and $\overline{F}=2.40 \times 10^{-3}$. Numerical results, shaded area and circles; Gaussian, black solid lines; skewed Gaussian, blue dashed lines; and saturation point, horizontal line. In panels (c) and (d), the symbols are numerical results and the solid lines are guides for the eye.  $L=18$, $\widehat{\mathcal{S}}^z=-3$, and ${\mathcal{D}} = 18\,564$; $J=1$ sets the energy scale.
}
\label{fig:skewed}
\end{figure}

In Fig.~\ref{fig:skewed}(a), we approximate the LDOS of an initial state with $E_{\text{ini}}$ far from the middle of the spectrum with a skewed Gaussian~\cite{Azzalini1985},
\begin{align}
P^{\text{ini}}_{\text{sG}}(E) &= 
\dfrac{2}{ \sqrt{ 2 \pi \sigma^2_{\text{s}} }  } \exp \left[ -\dfrac{(E-E_{\text{s}})^2}{ 2 \sigma^2_{\text{s}} }   \right] \Phi\left[\dfrac{\beta (E-E_{\text{s}})}{ \sigma_{\text{s}}} \right] , &\nonumber \\
&\Rightarrow F_{\text{sG}}(t)=4\exp (-\sigma^2_{\text{s}} t^2) \left| \Phi \left( \dfrac{i\beta\sigma_{\text{s}} t}{\sqrt{1+\beta^2} }\right) \right|^2,&
\end{align}
where
\[
\Phi(x)  = \dfrac{1}{2}  \left[ 1+ \text{erf} \left( \dfrac{x}{\sqrt{2 \sigma^2_{\text{s}}}} \right) \right] .
\]
Above, erf is the error function, $E_s$ and $\sigma_s$ are related to $E_\text{ini}$ and $\sigma_\text{ini}$ as  
\begin{equation*}
E_\text{s}=E_\text{ini}-\sigma_\text{s}\frac{\beta}{\sqrt{1+\beta^2}}\sqrt{\frac{2}{\pi}},
\end{equation*}
\begin{equation*}
\sigma_\text{s}=\sigma_\text{ini}\left(1-\frac{2}{\pi}\frac{\beta^2}{1+\beta^2}\right)^{-1/2},
\end{equation*}
and $\beta$ is related to the skewness $\gamma_1$ of the distribution,
\begin{equation}
\gamma_1 = \frac{\mu_3}{\sigma^3_{\text{ini}}}, \hspace{1 cm}
\mu_3= \sum_{\alpha} |C_{\alpha}^{\text{ini}} |^2 (E_{\alpha} - E_{\text{ini}})^3 ,
\label{eq:skewness}
\end{equation}
as 
\[
\gamma_1 = \frac{4-\pi}{2} \left[\frac{2\beta^2}{\beta^2 (\pi -2) + \pi} \right]^{3/2}.
\]
$P^{\text{ini}}_{\text{sG}}(E)$ is visibly a better match to the LDOS in Fig.~\ref{fig:skewed}(a) than the symmetric Gaussian. The corresponding fidelity curve for the skewed Gaussian is shown in Fig.~\ref{fig:skewed}(b). It is slower than what is obtained with a symmetric and well-filled Gaussian and closer to the actual numerical data.

To better quantify how much $P^{\text{ini}}(E)$ deviates from a Gaussian distribution as $E_{\text{ini}}$ moves away from the center of the spectrum, we show in Fig.~\ref{fig:skewed}(c) the skewness and in Fig.~\ref{fig:skewed}(d) the excess kurtosis,
\begin{equation}
\gamma_2 = \frac{\mu_4}{\sigma^4_{\text{ini}}} - 3, \hspace{1 cm}
\mu_4= \sum_{\alpha} |C_{\alpha}^{\text{ini}} |^2 (E_{\alpha} - E_{\text{ini}})^4 ,
\label{eq:kurtosis}
\end{equation}
of the LDOS for different values of $E_{\text{ini}}$. Notice that, just like $\sigma^2_{\text{ini}}$ in Eq.~(\ref{deltaE}), $\gamma_1$ and $\gamma_2$ can in principle be obtained before the diagonalization of the final Hamiltonian by computing the terms $\langle n|\widehat{H}_F|n'\rangle$.

For a skewed Gaussian function, the maximum values of the skewness and excess kurtosis are $\gamma_1=0.995$ and $\gamma_2=0.869$ \cite{Azzalini1985}. In Figs.~\ref{fig:skewed}(c) and~\ref{fig:skewed}(d), very far from the center of the spectrum, these values are larger than 1, indicating that the function that best represents $P^{\text{ini}}(E)$ in that region, at least for our system sizes, is probably not a skewed Gaussian, but some other skewed function. As $E_{\text{ini}}$ approaches the middle of the spectrum, the skewness and the excess kurtosis approach zero, as expected for a normal distribution.

%%%%%%%%%%%%%%%% SEMICIRCLE %%%%%%%%%%%%%%%%%%%%%%%%%
\section{AS FAST AS FULL RANDOM MATRICES}
\label{Sec:FRM}

Full random matrices are matrices completely filled with random numbers, where the only constraint is to satisfy the symmetries of the system to be studied. GOEs, for instance, address time-reversal invariant systems with rotational symmetry~\cite{Guhr1998}. These matrices describe well the statistical fluctuations of the spectrum, but they are unrealistic, because they imply the simultaneous interactions of many particles, while physical systems have few-body interactions. Full random matrices do not take into account the physical nature of the potential.

\begin{figure}[ht]
\centering
\includegraphics*[width=3.3in]{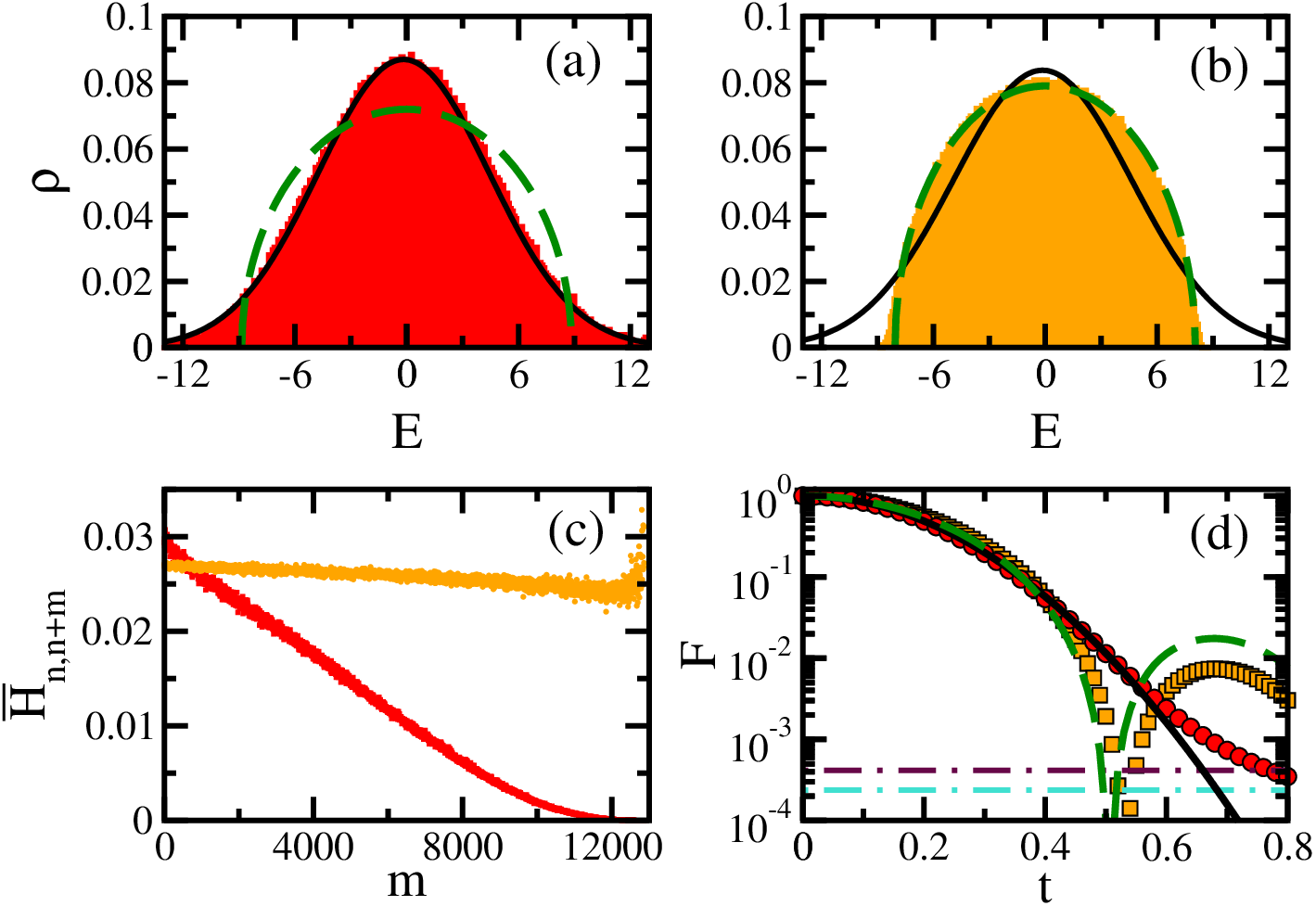}
\caption{(Color online) Density of states $\rho$ for the XXZ model with $\Delta_F=0.5$ and added random flip-flop terms between sites $i$ and $j$, where $j-1\geq 2$ (a), and with those elements replaced with uncorrelated random numbers (b). Average of the absolute value of the off-diagonal elements vs the distance $m$ from the diagonal (c) for the Hamiltonian from system (a) (decaying curve) and from system (b) (flat curve). Fidelity decay (d) for system (a) (Gaussian analytical expression is the black solid line; numerical data are the circles; saturation point is the highest horizontal line) and for system (b) (analytical expression is the green dashed line; numerical data are the squares; saturation point is the lowest horizontal line).  $\Delta_I=0.5$, $E_{\text{ini}} \sim 0$, $\sigma_{\text{ini}}=4.42$ (a), $\sigma_{\text{ini}}=4.03$ (b), $L=16$, $\widehat{\mathcal{S}}^z=0$, and ${\mathcal{D}} = 12\,870$; $J=1$ sets the energy scale.
}
\label{fig:random}
\end{figure}

Starting with the more realistic XXZ model, we studied the conditions under which the density of states would approach the semicircular shape of full random matrices by gradually adding random couplings between more and more distant pairs of spins and also between more than only two sites. Our Hamiltonian matrix was written in the site basis, that is product vectors where each site has a spin pointing either up or down in the $z$ direction. By including only flip-flop terms between distant pairs of spins, $J_{ij} (\widehat{S}_i^x \widehat{S}_{j}^x + \widehat{S}_i^y \widehat{S}_{j}^y)$, with $j-i\geq 2$ and $J_{ij} $ being random numbers from a Gaussian distribution with variance 1, the shape of the density of states remained Gaussian [Fig.~\ref{fig:random}(a)]. This was expected, since the system still had only two-body interactions; the matrix was sparse and its elements were correlated. However, the inclusion of hoppings involving four sites and of interactions of the kind $J_{ijk\ldots} \widehat{S}_{i}^z \widehat{S}_{j}^z \widehat{S}_{k}^z \ldots$ did not bring us any closer to a noticeable semicircle. Correlations seemed to be playing a major role. 

We then turned our attention back to the XXZ model where only flip-flop terms between any two sites were included, but now substituted the matrix elements corresponding to these couplings with uncorrelated random elements. Quite unexpectedly, because the matrix looked extremely sparse, a density of states very close to semicircular emerged [Fig.~\ref{fig:random}(b)].  This can be understood by analyzing the basis. The matrix is sparse in the site basis, but nearly full in the mean-field basis, that is, the basis corresponding to the eigenstates of the integrable part (XXZ) of the Hamiltonian. As seen in the plot for the averages of the absolute values of the off-diagonal elements,  
\begin{equation}
\overline{H}_{n,n+m}  =\frac{\sum_{n=1}^{{\cal D}-m} | H_{n,n+m} |}{{\cal D}-m},
\end{equation}
versus the distance $m$ from the diagonal [Fig.~\ref{fig:random}(c)], the matrix with uncorrelated randomized elements written in the mean-field basis is indeed filled with nonzero elements of similar amplitudes. In contrast, the off-diagonal elements of the Hamiltonian with random flip-flop terms decrease with $m$. This explains the different shapes of the density of states.
The form of the LDOS of initial states close to the middle of the spectrum of $\widehat{H}_F$ is similar to that of the density of states (not shown). The corresponding fidelity behaviors are shown in Fig.~\ref{fig:random} (d). It is Gaussian up to times close to $t_R$ for the Hamiltonian with random flip-flop terms and it is similar to $F_{\text{SC}}(t)$ [Eq.~(\ref{SemiCirc_SF})] for the Hamiltonian with uncorrelated elements. The agreement with $F_{\text{SC}}(t)$ becomes even better if uncorrelated random elements replace also matrix elements associated with flip-flop terms involving four sites (not shown).

In Sec.~\ref{sec:transition}, we saw that, in a system with a Gaussian density of states, the increase of the perturbation strength broadens the local density of states from Lorentzian to Gaussian. In the present section, we provided a simple recipe to achieve the transition from a Gaussian to a semicircle density of states, which causes  the same change in the local density of states. The transition of the shape of the local density of states from Lorentzian to Gaussian and then finally to semicircle was investigated before~\cite{Fyodorov1996} in the context of band random matrices. The latter were introduced by Wigner~\cite{Wigner1955,Wigner1957} in an attempt to improve over the unrealistic scenario of full random matrices. An important advantage of our analysis over (band or full) random matrices is to address realistic models associated with a very broad range of physical systems.

%%%%%%%%%%%%%%%% COS SQUARED %%%%%%%%%%%%%%%%%%%%%%%%%
\section{ABSOLUTE LOWER BOUND}
\label{Sec:Cos}

In Ref.~\cite{Torres2014PRE}, we studied the case of a local quench in space, where we added to $\widehat{H}_I=\widehat{H}_\text{NN}$ a static magnetic field localized on site $L/2$ and leading to an excess energy of amplitude $d_F$. For $d_F\lesssim1$, we verified that $P^{\text{ini}}(E)$ could not reach the Gaussian shape observed for global quenches (that is, perturbations affecting all the sites of the chain, as the quench to the NNN model) and was instead restricted to the Lorentzian form. In the current article, we analyze local quenches where $d_F>1$ and the chain effectively splits in two. 
\begin{figure}[htb]
\centering
\includegraphics*[width=3.3in]{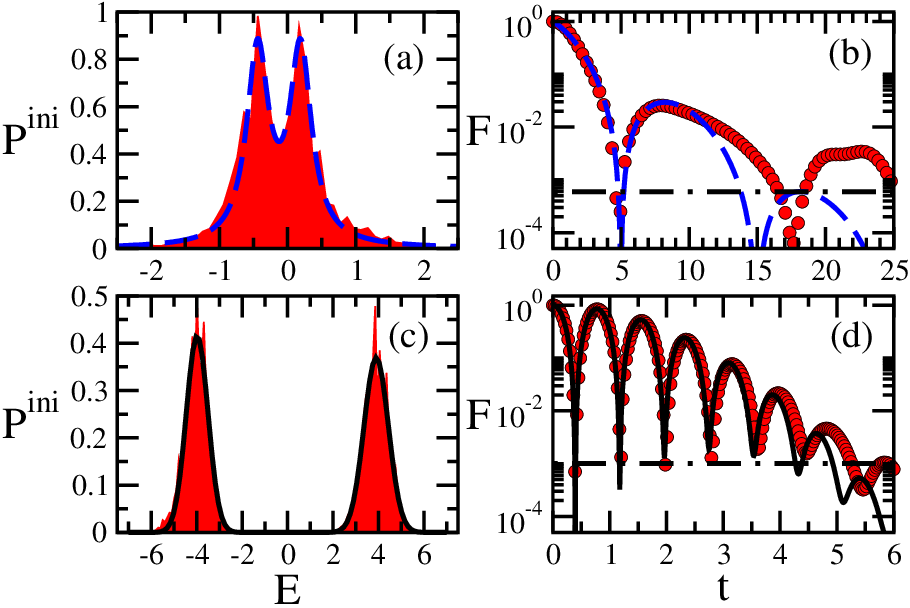}
\caption{(Color online) Local density of states (a, c) and corresponding fidelity decay (b, d) for a quench from the XXZ model to the impurity model with $d_F=1.2$ (a, b) and $d_F=8.0$ (c, d). The initial state is in the middle of the spectrum of $\widehat{H}_I$, ${\cal E}_{{\cal D}/2}$. (a, c) Numerical data, shaded red area; two Lorentzians  with $E_1=-0.44$, $E_2=0.19$, and $\Gamma_1=\Gamma_2=0.39$, blue dashed line (a); two Gaussians with $E_1=-3.98$, $E_2=3.90$, $\sigma_1=0.48$, and $\sigma_2=0.54$, black solid line (c). (b, d) Numerical results (circles); Eq.~(\ref{eq:TL}), blue dashed line; and Fourier transform of the two Gaussians from panel (c), black solid lines. The saturation points are the horizontal lines. $\Delta_I=\Delta_F=0.48$, $L=16$, $\widehat{\mathcal{S}}^z=0$, and ${\mathcal{D}} = 12\,870$; $J=1$ sets the energy scale.
}
\label{fig:bimodal}
\end{figure}

As $d_F$ increases, the density of states of $\widehat{H}_F$ eventually divides in two peaks. The crossover from a unimodal to a bimodal distribution is carried on also to $P^{\text{ini}}(E)$, where $|\text{ini}\rangle$ is an  eigenstate of the XXZ model. When $d_F \gtrsim 1$, the single Lorentzian for $P^{\text{ini}}(E)$ starts splitting in two Lorentzians. For the ${\cal{S}}^z=0$ sector and $E_{\text{ini}}$ close to the middle of the spectrum, both equally weighted $P_L(E)$, one centered at $E_1$ and the other at $E_2$, have approximately the same width, as shown in Fig.~\ref{fig:bimodal}(a). They lead to
\begin{equation}
F_{\text{TL}}(t) =\cos^2\left(\frac{E_2 - E_1}{2} t\right)\exp(-\Gamma t),
\label{eq:TL}
\end{equation}
where $E_2-E_1 \approx d_F$. The expression above matches well the numerical data of the fidelity decay for a fairly long time in Fig.~\ref{fig:bimodal}(b). The oscillations after $t_R$ are exponentially suppressed with a rate determined by the width of the Lorentzians.

As $d_F$ further increases, 
the peaks broaden and approach Gaussians separated in energy by $E_2-E_1\approx d_F$.  When both peaks have the same width $\sigma$,
\begin{equation}
F_{\text{TG}}(t) = \cos^2\left(\frac{E_2 - E_1}{2} t\right)\exp(-\sigma^2 t^2).
\label{eq:TG}
\end{equation}
The envelope of the decaying oscillations of the fidelity is now also Gaussian. 
This scenario is illustrated in Fig.~\ref{fig:bimodal}(c), although the widths of the Gaussians there are slightly different. As shown in Fig.~\ref{fig:bimodal}(d), the corresponding Fourier transform of the two Gaussians agrees very well with the numerical results for the fidelity decay until very close to its saturation.

As the energy of the initial state moves away from the center of the spectrum, $P^{\text{ini}}(E)$ becomes,  as expected, more asymmetric. Larger contributions to the distribution appear for the peak closer to the border of the spectrum. The fidelity decay becomes slower if compared to states where $E_{\text{ini}}$ is closer to the middle of the spectrum. However, for very large $d_F$, the asymmetry decreases and both peaks approach Gaussians.

From Eqs.~(\ref{eq:TL}) and (\ref{eq:TG}), one sees that for $t<\pi/(2\sigma_{\text{ini}})$, where the total dispersion in energy $\sigma_{\text{ini}} \approx  (E_2 - E_1)/2 \approx d_F/2$, the fidelity decay derived from bimodal distributions can indeed approach the ultimate bound associated with the energy-time uncertainty relation, $F \geq \cos^2 (\sigma_{\text{ini}} t)$. This is particularly evident when $d_F$ is large, since in this case $\sigma^2 t^2 < \sigma^2 \pi^2/(E_2 - E_1)^2 \ll 1$ and we obtain 
\begin{equation}\label{eq:ultimate}
F_{\text{TG}}(t) \sim  \cos^2 \left(\dfrac{d_F t}{2 }\right).
\end{equation}

The lower bound for the fidelity decay can be obtained from the Mandelstam-Tamm uncertainty relation~\cite{Mandelstam1945}
\[
\sigma_{\text{ini}} \sigma_A \geq \frac{1}{2} \left| \frac{d\langle \hat{A} \rangle}{dt} \right| .
\]
A pedagogical derivation is provided in Ref.~\cite{Uffink1993}. If $\hat{A}$ is the projection operator on the initial state, $\hat{A}=|\text{ini}\rangle \langle \text{ini}|$, then 
$\langle \hat{A} \rangle = F(t)$ and $\sigma_A^2 = F(t) - F(t)^2$. Thus
\[
\sigma_{\text{ini}} \sqrt{F(1-F)} \geq \frac{1}{2} \left| \frac{dF}{dt} \right|,
\]
which leads to
\[
 \arccos \left( \sqrt{F(t)} \right)  \geq \sigma_{\text{ini}} t \Rightarrow F(t) \geq \cos^2(\sigma_{\text{ini}} t).
 \]

\section{CONCLUSION}
\label{Sec:Conclusion}

We studied isolated finite quantum systems described by one-dimensional spin-1/2 models with two-body interactions and taken far from equilibrium instantaneously. We denominated as fidelity the probability for finding the initial state later in time. This probability corresponds to the Fourier transform of the weighted energy distribution of the initial state (LDOS), $P^{\text{ini}}(E)$. We analyzed three realistic scenarios in which the fidelity decay was nonexponential. The first two cases involved global quenches in space, where $P^{\text{ini}}(E)$ was unimodal. The third one occurred after a local quench, which resulted in a bimodal $P^{\text{ini}}(E)$.

{\em Case (1)}. When $E_{\text{ini}}$ is close to the middle of the spectrum of the postquench Hamiltonian and the global perturbation is strong, $P^{\text{ini}}(E)$ is Gaussian, leading to a Gaussian fidelity decay. This behavior is independent of the regime (integrable or chaotic) of the system. It can hold for long times and even persist up to saturation. 

Before reaching the Gaussian regime, as the perturbation increases, $P^{\text{ini}}(E)$ goes first from a Lorentzian shape to a convolution between Lorentzian and Gaussian (Voigt distribution).  Equivalently, the fidelity decay mixes Gaussian and exponential functions. 

In the limit of strong perturbation, but far from the middle of the spectrum, where there are fewer states and finite effects are important, $P^{\text{ini}}(E)$ can be approximated by a skewed Gaussian. In this case, the fidelity decay is slower than Gaussian. 

{\em Case (2)}. The fidelity decay is faster than Gaussian and approaches the results from full random matrices if the matrix elements of the spin-1/2 Hamiltonian corresponding to long-range flip-flop terms are replaced with uncorrelated random numbers. This matrix is very sparse in the site basis, but it is nearly filled if written in the mean-field basis.

{\em Case (3)}. After a local quench, where a strong static magnetic field is added to a single site of the chain, $P^{\text{ini}}(E)$ becomes bimodal. The initial fidelity decay is approximately the one established by the energy-time uncertainty relation. After crossing the saturation point for the first time, the envelope of the subsequent oscillations decays as an exponential or Gaussian, depending on the shape of the two peaks in $P^{\text{ini}}(E)$. 

In Refs.~\cite{Torres2014PRA,Torres2014NJP}, we discussed initial states accessible to experiments with optical lattices, where Case (1) could be tested. The local quench described in Case (3) is also viable to those experiments. Another important aspect of the present work is the connection between fidelity decay and studies in spectroscopy.
The tools used for determining lifetime and line shape in that field can be very useful in the analysis of quench dynamics.

%%%%%%%%%%%%%%%%%%%% ACKNOWLEDGMENTS %%%%%%%%%%%%%%%%%%%%%
\vspace{-0.2cm}
\section*{ACKNOWLEDGMENTS}
This work was supported by the NSF (USA) under Grant~No. DMR-1147430 and partially under Grant No. PHYS-1066293. L.F.S. thanks the hospitality of the Aspen Center for Physics. E.J.T.H. acknowledges partial support from  \mbox{CONACyT}, Mexico. We thank F. M. Izrailev for the careful reading of the manuscript and useful suggestions.
\vspace{-0.4cm}

%%%%%%%%%%%%%%%%%%%% REFERENCES %%%%%%%%%%%%%%%%%%%%%

\end{document}